\documentclass[runningheads,a4paper]{llncs}

\usepackage{epsfig}
\usepackage{amsmath}
\usepackage{multirow}
\usepackage{amssymb}
\usepackage{graphicx}
\usepackage{array}

\usepackage{url}
\urldef{\mails}\path|{lanbo, kailu, manazhao, yiz}@soe.ucsc.edu|    
\newcommand{\keywords}[1]{\par\addvspace\baselineskip
\noindent\keywordname\enspace\ignorespaces#1}

\newcommand{\set}[1]{\mathbf{#1}}

\begin{document}

\mainmatter  

\title{Interactive Retrieval Based on Wikipedia Concepts}


%
%
\author{
Lanbo Zhang
}
%

\institute{
Twitter Inc. \\
lanboz@twitter.com \\
}

%
%

\maketitle

\begin{abstract}
This paper presents a new user feedback mechanism based on Wikipedia concepts for interactive retrieval. In this mechanism, the system presents to the user a group of Wikipedia concepts, and the user can choose those relevant to refine his/her query. To realize this mechanism, we propose methods to address two problems: 1) how to select a small number of possibly relevant Wikipedia concepts to show the user, and 2) how to re-rank retrieved documents given the user-identified Wikipedia concepts. Our methods are evaluated on three TREC data sets. The experiment results show that our methods can dramatically improve retrieval performances.
\end{abstract}

\section{Introduction}

As an online encyclopedia, Wikipedia has covered a wide range of subjects and millions of concepts. Ad-hoc retrieval queries often relate to some concepts, which mostly have been covered by Wikipedia. For example, the query ``Ford foreign ventures'' relates to the Wikipedia concepts ``Ford Motor Company'', ``Joint Venture'', etc., the query ``Economic espionage'' relates to the Wikipedia concepts ``Industrial Espionage'', ``Trade Secret'', etc. Since Wikipedia contains a lot of valuable and high-quality information, if we can identify the related Wikipedia concepts of a query, we might be able to leverage the corresponding Wikipedia information to facilitate document retrieval. To achieve this goal, this paper presents a new user feedback mechanism based on Wikipedia concepts. In this mechanism, whenever the user inputs a query, the system will show a group of Wikipedia concepts, and the user can choose those relevant ones to refine his/her information need. To realize this mechanism, we need to answer two questions: 1) how to select a small number of possibly relevant concepts given the user query? and 2) how to rank documents when the user-identified concepts are available? To address these two questions, we propose several methods based on different sources of evidence. The proposed methods are evaluated on three data sets from TREC along with a user study on Mechanical Turk. Our experiment results show that using user-identified Wikipedia concepts can significantly improve retrieval performances on all three data sets.

\section{Related Work}

Wikipedia has been shown to be a useful resource for many intelligent tasks, including pseudo relevance feedback \cite{xu2009query}, query expansion \cite{li2007improving}, cross-language information retrieval \cite{schonhofen2007performing,potthast2008wikipedia}, text classification \cite{gabrilovich2007harnessing}, etc. Xu et al. \cite{xu2009query} propose a query-dependent method for selecting Wikipedia articles for pseudo relevance feedback. Li et al. \cite{li2007improving} propose to use Wikipedia as an external corpus to expand difficult queries. In this paper, we explore the usage of Wikipedia in interactive retrieval and propose a new user feedback mechanism based on Wikipedia.

Different types of user feedback have been shown useful for ad-hoc retrieval, including document-based relevance feedback \cite{rocchio1971relevance,salton1997improving,Xing2011bias}, term-based feedback \cite{tan2007term}, metadata-based faceted feedback \cite{zhang2010interactive,zhang2011filtering,Zhang:2013:CFS:2604906,zhang2010discriminative}. In this paper, we study a new type of user feedback based on Wikipedia concepts and show that this type of feedback can be very useful for retrieval.

Query expansion is a fundamental technique for dealing with the term mismatch problem in information retrieval. The basic idea is to find additional terms that are related to the underlying information need to expand the user query. Many methods for term selection have been studied \cite{xu1996query,voorhees1994query,qiu1993concept,buckley1995automatic,mitra1998improving,robertson1990term,zhang2009ucsc}. In this paper, our document ranking methods are based on query expansion. We rely on user-identified Wikipedia concepts to select high-quality terms for query expansion.

\section{User Feedback Mechanism Based on Wikipedia Concepts}

\begin{figure}
\centering
\epsfig{file=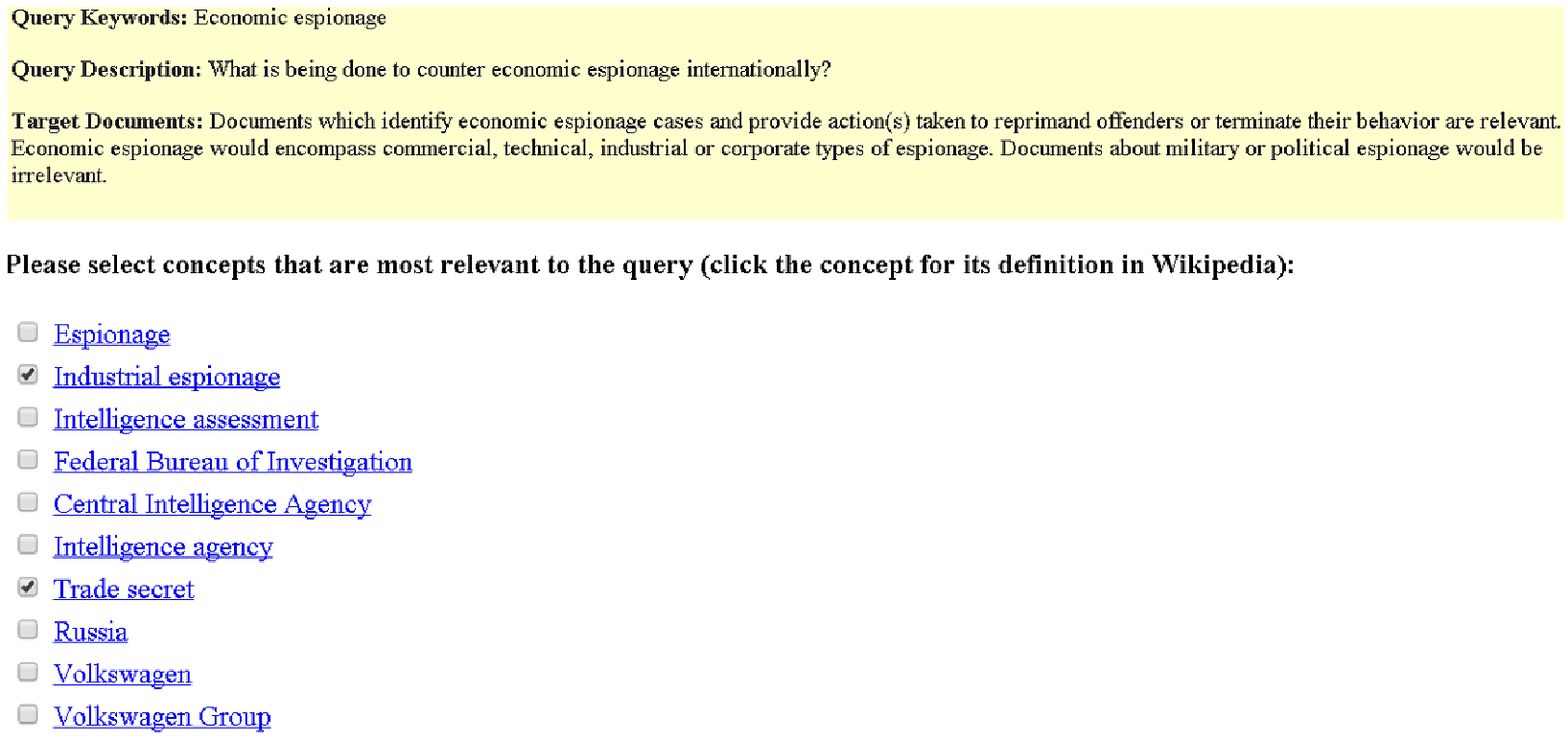, scale=0.7}
\caption{User interface for collecting user feedback on Wikipedia concepts}\label{fig:mturk}
\end{figure}

The user feedback mechanism based on Wikipedia concepts contains four interactive steps: 1) the user inputs the query; 2) the system shows a small number of possibly relevant Wikipedia concepts; 3) the user identifies the relevant concepts; 4) the system re-ranks the retrieved documents. Figure \ref{fig:mturk} shows the interface for collecting user feedback on Wikipedia concepts in our user study, where each concept is represented by the title of the corresponding Wikipedia article, and the user can click on a concept to navigate to the Wikipedia page.

Compared with other types of user feedback including document-based and term-based feedback, the new type of feedback based on Wikipedia concepts has several advantages. First, each concept is represented by a short title, thus it will take the user much less time to review a group of concepts than reviewing a group of documents. Second, unlike individual terms, which might be ambiguous without a context, each Wikipedia concept has a clear semantic meaning and is easy for users to understand. The user can always navigate to the Wikipedia page for detailed explanation if he/she is not familiar with the concept. Third, there are relevant Wikipedia concepts for most queries because of the comprehensive coverage of Wikipedia. Even for difficult queries for which existing retrieval methods fail to find relevant documents in the corpus, it's possible to find Wikipedia concepts that are relevant to the query and thus useful for retrieval.

To realize the new user feedback mechanism, there are two major research questions we need to answer: 1) how to select a small number of Wikipedia concepts that are very likely to be relevant to the query? 2) how to re-rank the retrieved documents given the user-identified relevant concepts? We will address these two questions in the following two sections.

\section{Wikipedia Concept Selection}

Users are generally not willing to spend much time reviewing a lot of concepts, thus it's important to select a small number of candidates which are very likely to be relevant. We propose several methods that are based on different sources of evidence for measuring the relevance degree between a concept and the query.

\subsection{Concept Occurrences in Top Ranked Documents}

This method scores each concept based on its occurrences in the top ranked documents. We first rank all documents in the corpus based on the initial query, then count the occurrences of each concept in the top N documents. Specifically, each concept is scored by the following Equation:

\begin{equation}
score_{\textbf{TD}}(c)=\sum_{d\in \set{D}}{w(c,d)*rank(d)^{-\beta}}
\end{equation}

where $c$ is a Wikipedia concept, $d$ is a document, $\set{D}$ is the set of all documents in the corpus, $w(c,d)$ is the weight of concept $c$ in document $d$, $rank(d)$ is the rank of $d$ in the initial document ranking based on the query (starting with 1), $\beta$ is a parameter that controls the decreasing speed of document weight along with its ranking position. We use a Wikification tool \cite{wikifier} to annotate all documents in the corpus, thus $w(c,d)$ can be calculated using the BM25 weighting function as follows,

\begin{equation}\label{documentconceptweighting}
w(c,d)=\frac{freq(c, d)}{freq(c, d) + 0.5 + 1.5 * \frac{length(d)}{avgDocLength}}
\end{equation}

where $freq(c,d)$ is the number of times that concept $c$ is annotated in document $d$.

This scoring method is based on the following intuition: the more frequently a concept occurs in top documents, the more likely it is relevant to the query. Documents that are ranked higher are more likely to be relevant, thus the factor $rank(d)^{-\beta}$ is used to capture the decrease of document relevance probability along with its ranking position.

\subsection{Concept Title}

The relevance degree between the query and a concept can be measured by the term match between the query and the concept. This method scores each concept by the following Equation:

\begin{equation}\label{equConceptScoring}
score(c)=\sum_{t\in q}{w(t,q) * w(t,c) * IDF(t)}
\end{equation}

where $w(t,q)$ is the term weight in the query, which is calculated by the following equation:

\begin{equation}
w(t,q) = \frac{tf(t, q)}{tf(t, q) + 2.0}
\end{equation}

$w(t,c)$ is the term weight of $t$ associated with concept $c$. There are multiple ways to calculate $w(t,c)$. The simplest way is to use the concept title, based on which we can calculate $w(t,c)$ as follows,

\begin{equation}\label{cttermweight}
w_{\textbf{CT}}(t,c) = \frac{tf(t, CT_c)}{tf(t, CT_c) + 0.5 + 1.5 * \frac{length(CT_c)}{avgConceptTitleLength}}
\end{equation}

$CT_c$ is the concept title of $c$. For example, the concept ``United States'' has two terms (``United'' and ``States'') in its title. $tf(t,CT_c)$ is the term frequency of $t$ in the concept title. We choose to use the BM25 weighting function since it has been shown to be more effective in measuring short text similarity \cite{zhang2012summarizing}.

\subsection{Wikipedia Article}

This method calculates $w(t,c)$ based on the Wikipedia article. Let $WA_c$ be the corresponding Wikipedia article of concept $c$, the term weight associated with $c$ can be calculated as follows,

\begin{equation}\label{watermweight}
w_{\textbf{WA}}(t,c)= \frac{tf(t, WA_c)}{tf(t, WA_c) + 0.5 + 1.5 * \frac{length(WA_c)}{avgWikiDocLength}}
\end{equation}

Equation \ref{equConceptScoring} is then used to score each concept.

\subsection{Anchor Texts}

Anchor texts are those words/expressions that are linked to the Wikipedia concept by the Wikification tool \cite{wikifier}. Let $\set{A}_c$ be the set of all anchor texts of concept $c$, $w(t,c)$ can be calculated as follows,

\begin{equation}\label{attermweight}
w_{\textbf{AT}}(t,c)=\sum_{a\in \set{A}_c}{\frac{tf(t, a)}{tf(t, a) + 0.5 + 1.5 * \frac{length(a)}{avgAnchorTextLength}}}
\end{equation}

where $a$ is a particular anchor text of $c$.

\subsection{Related Documents}

This method calculates $w(t,c)$ based on the related documents which are annotated with the concept $c$. The following equation is used for term weighting:

\begin{equation}\label{rdtermweight}
w_{\textbf{RD}}(t,c)=\sum_{d\in {\set{D}}}{w(t,d) * w(c,d)}
\end{equation}

where $D$ is the set of all documents in the corpus, $w(c,d)$ is the weight of concept $c$ in document $d$, which is calculated using Equation \ref{documentconceptweighting}, and $w(t,d)$ is the weight of term $t$ in document $d$, which is calculated using the BM25 weighting function similar to Equation \ref{documentconceptweighting}.

\subsection{Overall Concept Scoring Function}

All above introduced methods are combined for concept selection. The overall concept scoring function is as follows,

\begin{equation}\label{conceptscoringfunction}
\begin{split}
score(c)=&N(score_{\textbf{TD}}(c)) \\
 + \alpha_1 * &N(score_{\textbf{CT}}(c)) \\
 + \alpha_2 * &N(score_{\textbf{WA}}(c)) \\
 + \alpha_3 * &N(score_{\textbf{AT}}(c)) \\
 + \alpha_4 * &N(score_{\textbf{RD}}(c)) \\
\end{split}
\end{equation}

where $\alpha$'s are parameters we can set or tune, $N(*)$ is the score normalization function, where the standard normalization\footnote{Standard normalization is to transform scores so that they have a mean of zero and a standard deviation of one} is used.

\section{Document Re-Ranking with User Feedback on Wikipedia Concepts}\label{sec:docranking}

After receiving user feedback on Wikipedia concepts, the system needs to re-rank documents in the corpus. The system now has access to the initial query as well as a group of user-identified Wikipedia concepts. We propose several methods for document ranking that are based on different sources of evidence. A synthesized ranker that combines all ranking methods is then used for document ranking.

\subsection{Concept Match}

This method scores each document based on the overlapping of user-selected concepts and those annotated with the document. Let $\set{C}_u$ be the set of all user-selected concepts, a document is scored as follows,

\begin{equation}
score_{\textbf{CM}}(d) = \sum_{c\in \set{C}_u}{w(c,d)}
\end{equation}

where $w(c,d)$ is the weight of concept $c$ in document $d$ by Equation \ref{documentconceptweighting}.

\subsection{Concept Titles}

In this method, the user-selected concepts are expanded with related terms. A relevance model in the form of a group of weighted terms will be calculated based on user-selected concepts. Let $w(t,\set{C}_u)$ be the weight of term $t$ in the relevance model, each document will be scored by the following Equation:

\begin{equation}
score(d) = \sum_{t\in d}{w(t,\set{C}_u) * w(t,d) * IDF(t)}
\end{equation}

To calculate the relevance model $w(t,\set{C}_u)$, we can use different sources of evidence as used for concept selection. Based on the concept title, the relevance model can be calculated as follows,

\begin{equation}
w_{\textbf{CT}}(t,\set{C}_u) = \sum_{c\in \set{C}_u}{w_{\textbf{CT}}(t,c)}
\end{equation}

where $c$ is one of the user-selected concepts, $w_{\textbf{CT}}(t,c)$ is the weight of $t$ in the title of $c$, which is calculated using Equation \ref{cttermweight}.

\subsection{Wikipedia Articles}

This method calculates the relevance model based on the corresponding Wikipedia articles of user-selected concepts.

\begin{equation}
w_{\textbf{WA}}(t,\set{C}_u) = \sum_{c\in \set{C}_u}{w_{\textbf{WA}}(t,c)}
\end{equation}

where $w_{\textbf{WA}}(t,c)$ is the weight of $t$ in the corresponding Wikipedia article of $c$, which is calculated using Equation \ref{watermweight}. In our experiments, only the top 20 terms with the largest values of $w_{WA}(t,\set{C}_u) * IDF(t)$ are kept and used in the relevance model.

\subsection{Anchor Texts}

This method calculates the relevance model based on anchor texts of user-selected concepts.

\begin{equation}
w_{\textbf{AT}}(t, \set{C}_u) = \sum_{c\in \set{C}_u}{w_{\textbf{AT}}(t,c)}
\end{equation}

where $w_{\textbf{AT}}(t,c)$ is the weight of $t$ in the corresponding anchor texts of $c$ , which is calculated using Equation \ref{attermweight}.

\subsection{Related Documents}

This method calculates the relevance model based on the related documents of user-selected concepts. Related documents are those annotated with the concept.

\begin{equation}
w_{\textbf{RD}}(t, \set{C}_u) = \sum_{c\in \set{C}_u}{w_{\textbf{RD}}(t,c)}
\end{equation}

where $w_{\textbf{RD}}(t,c)$ is the weight of $t$ in related documents of $c$, which is calculated using Equation \ref{rdtermweight}.

\subsection{Overall Document Ranking Function}

We combine all above methods for document ranking. The overall document scoring function is as follows,

\begin{equation}\label{docrankingfunc}
\begin{split}
score(d) = &N(score_{\textbf{IQ}}(d)) \\
 + \beta_1 * &N(score_{\textbf{CM}}(d)) \\
 + \beta_2 * &N(score_{\textbf{CT}}(d)) \\
 + \beta_3 * &N(score_{\textbf{WA}}(d))\\
 + \beta_4 * &N(score_{\textbf{AT}}(d)) \\
 + \beta_5 * &N(score_{\textbf{RD}}(d))\\
\end{split}
\end{equation}

where $score_{\textbf{IQ}}(d)$ is the document score based on the initial query, $score_{\textbf{CM}}(d)$ is the document score based on concept match between user-selected and document-annotated concepts, the other scores are based on term match between the document and different types of information: concept titles (CT), Wikipedia articles (WA), anchor texts (AT), and related documents (RD) respectively, $N(*)$ is the standard normalization function.

\section{Experimental Methodology}

\subsection{Data Sets}
Three TREC data sets are used to evaluate the proposed methods. \textbf{Filtering-02} consists of the RCV1 corpus and 50 topics from the TREC 2002 filtering track. RCV1 contains a total number of 806791 news articles. The 50 topics created by human assessors are used as queries \cite{robertson2002trec}. \textbf{HARD-03} consists of the AQUAINT corpus and 50 topics from the TREC 2003 HARD track \cite{allan2005hard}. AQUAINT contains approximately 1,033,000 news articles \cite{voorhees2005trec}. \textbf{HARD-05} consists of the AQUAINT corpus and 50 topics from the TREC 2005 HARD track.

To prepare the list of Wikipedia concepts for concept selection, we extract all concepts from the English Wikipedia dump of August 5, 2013. We also use the wikification tool developed by Roth and Ratinov \cite{wikifier} to annotate all documents in RCV1 and AQUAINT.





\subsection{User Feedback Collection}

We use Mechanical Turk to collect user feedback on Wikipedia concepts. For each query, we show workers information about the TREC topic including keywords, description, and narrative as well as 20 selected Wikipedia concepts. Workers are asked to identify concepts that are relevant to the query. Figure \ref{fig:mturk} shows the task interface on Mechanical Turk. We have 5 workers to work on each query, and the average performance will be reported. 

\subsection{Experimental Settings}

To evaluate the utility of user-identified Wikipedia concepts, we compare the retrieval performances of several runs. The baseline run only uses the initial query with the BM25 retrieval algorithm \cite{robertson1999okapi}. Other runs use both the initial query and user-selected Wikipedia concepts with document ranking methods proposed in Section \ref{sec:docranking} respectively.

To evaluate concept selection methods, a straightforward approach is to compare the retrieval performance each method leads to. However, this evaluation approach is very costly since we would need to collect a number of users' feedback for each of the concept selection methods. Instead, we prepare a large set of concepts by mixing concepts selected by all methods, and hire a human assessor to judge the relevance of each concept in this set. Each concept selection method is then used to rank concepts in the large set, and evaluated by how well its ranking matches the one by the human assessor, where the NDCG measure is used.

For parameter tuning of all methods, we use 2-fold cross validation by randomly splitting 50 topics in each data set into two folds.

\section{Experiment Results}

We try to answer two questions based on our experiment results: 1) Are the user-identified Wikipedia concepts useful for document retrieval? 2) How do the concept selection methods perform? Which one performs the best?

\subsection{Utility of User-Identified Wikipedia Concepts}

Table \ref{tab:retperf} compares the retrieval performances with and without using user-identified Wikipedia concepts. BM25 is used in the baseline run, Equation \ref{docrankingfunc} is used in the experiment run with user feedback. Note that the first item of Equation \ref{docrankingfunc} is equivalent to the baseline method. For parameter tuning of all methods, MAP is the measure we try to optimize.

According to Table \ref{tab:retperf}, MAP is significantly improved on all three data sets, P@10 is significantly improved on HARD-03 and HARD-05, while not significantly on Filtering-02, probably because P@10 of the baseline run is already high on this data set. Particularly, MAP and P@10 on HARD-05 are dramatically increased by around 50\%, which means user-identified Wikipedia concepts are very useful on this data set.

It's interesting to understand why relevant Wikipedia concepts can be helpful in retrieval. We analyzed quite a few queries for which the retrieval performance is dramatically improved, and found the following explanations. First, relevant Wikipedia concepts can serve as a good source of terms for query expansion. For example, for query ``mercy killing'', the relevant concept ``Euthanasia'' provides a good source of expanding terms. Second, relevant Wikipedia concepts can emphasize the important aspects of a query that are easy to be ignored in baseline retrieval. For example, for query ``Archaeology discoveries'', the relevant concepts ``Archaeology'' and ``Artifact (archaeology)'' will boost discoveries in Archaeology and inhibit discoveries in other subjects. Another example is query ``recycle, automobile tires'', for which the relevant concept ``Tire recycling'' helps filter out a lot of high-ranked documents about automobile recycling instead of tire-specific recycling. Third, relevant Wikipedia concepts can help reduce ambiguity in retrieval. For example, the query ``piracy'' has two meanings and the user-desired stories should be about ship-taking practices instead of copyright infringements, the relevant concept ``Piracy in Somalia'' will be very useful for disambiguation between these two meanings. Another example is query ``human smuggling'', for which the aspect ``human'' is easy to be ignored, and a lot of high-ranked documents are about smuggling of goods instead of humans. In this example, the relevant concept ``People smuggling'' will be very helpful.

\begin{table}[ht]
\caption{Retrieval performances without and with user-identified Wikipedia concepts. $^\ast$ indicates the improvement over baseline is statistically significant under the paired t-test with significance level 0.05.}\label{tab:retperf}
\begin{center}
\begin{tabular}{|p{3.6cm}|>{\centering\arraybackslash}m{1.2cm}|>{\centering\arraybackslash}m{1.2cm}|>{\centering\arraybackslash}m{1.2cm}|>{\centering\arraybackslash}m{1.2cm}|>{\centering\arraybackslash}m{1.2cm}|>{\centering\arraybackslash}m{1.2cm}|}
\hline
\bf{Data Set} & \multicolumn{2}{|c|}{\bf{Filtering-02}} & \multicolumn{2}{|c|}{\bf{HARD-03}} & \multicolumn{2}{|c|}{\bf{HARD-05}} \\
\hline
\bf{Measure} & \bf{MAP} & \bf{P@10} & \bf{MAP} & \bf{P@10} & \bf{MAP} & \bf{P@10} \\
\hline
Baseline (BM25) & 0.274 & 0.456 & 0.163 & 0.202 & 0.196 & 0.320\\
\hline							
With Feedback & 0.333 & 0.486 & 0.192 & 0.242 & 0.312 & 0.464\\
\hline
Improvement & 21.5\%$^\ast$ & 6.6\% & 17.8\%$^\ast$ & 19.8\%$^\ast$ & 59.2\%$^\ast$ & 45.0\%$^\ast$ \\
\hline
\end{tabular}
\end{center}
\end{table}

Table \ref{tab:featureperf} shows the retrieval performance of each document ranking method proposed in Section \ref{sec:docranking}. Among all individual methods, WA consistently performs the best on all data sets, which means Wikipedia articles is the best source of terms for query expansion. We also find that when more and more ranking methods are combined, the retrieval performances keep increasing, and the performances are best when all methods are combined. This implies that the proposed methods measure different aspects of document relevance and are complementary with each other.

\begin{table}[ht]
\caption{Retrieval performances of different document ranking methods.}\label{tab:featureperf}
\begin{center}
\begin{tabular}{|p{3.6cm}|>{\centering\arraybackslash}m{1.2cm}|>{\centering\arraybackslash}m{1.2cm}|>{\centering\arraybackslash}m{1.2cm}|>{\centering\arraybackslash}m{1.2cm}|>{\centering\arraybackslash}m{1.2cm}|>{\centering\arraybackslash}m{1.2cm}|}
\hline
\bf{Data Set} & \multicolumn{2}{|c|}{\bf{Filtering-02}} & \multicolumn{2}{|c|}{\bf{HARD-03}} & \multicolumn{2}{|c|}{\bf{HARD-05}} \\
\hline
\bf{Measure} & \bf{MAP} & \bf{P@10} & \bf{MAP} & \bf{P@10} & \bf{MAP} & \bf{P@10} \\
\hline
IQ (Baseline) & 0.274 & 0.456 & 0.163 & 0.202 & 0.196 & 0.320\\
\hline							
WA (Wikipedia Article) & 0.260 & 0.402 & 0.155 & 0.218 & 0.249 & 0.374\\
\hline
CM (Concept Match) & 0.195 & 0.338 & 0.082 & 0.144 & 0.135 & 0.270\\
\hline
CT (Concept Title) & 0.248 & 0.424 & 0.136 & 0.154 & 0.191 & 0.304\\
\hline
AT (Anchor Texts) & 0.236 & 0.380 & 0.120 & 0.158 & 0.206 & 0.302\\
\hline
RD (Related Docs) & 0.221 & 0.324 & 0.110 & 0.134 & 0.147 & 0.232\\
\hline
IQ+WA & 0.319 & 0.482 & 0.191 & 0.238 & 0.289 & 0.446\\
\hline
IQ+WA+CM & 0.321 & 0.478 & 0.192 & 0.242 & 0.300 & 0.460\\
\hline
IQ+WA+CM+CT & 0.328 & 0.472 & 0.193 & 0.230 & 0.301 & 0.448\\
\hline
IQ+WA+CM+CT+AT & 0.332 & 0.480 & 0.192 & 0.232 & 0.305 & 0.448\\
\hline
All & 0.333 & 0.486 & 0.192 & 0.242 & 0.312 & 0.464\\
\hline
\end{tabular}
\end{center}
\end{table}

\subsection{Wikipedia Concept Selection}

Table \ref{tab:conceptselectionperf} shows the performances of all concept selection methods. For each method, the NDCG is calculated based on how well each method's ranking of concepts corresponds to that of the human assessor. According to the table, it's clear that the synthesized method that combines all methods performs the best. This implies that the individual evidences of concept relevance are complementary with each other and can be combined to achieve a better performance. Besides, none of the individual methods consistently performs the best on all data sets, thus no conclusion can be made on which individual evidence is most useful.

\begin{table}[ht]
\caption{Concept selection performances (NDCG) of different methods.}\label{tab:conceptselectionperf}
\begin{center}
\begin{tabular}{|p{3.6cm}|>{\centering\arraybackslash}m{2.4cm}|>{\centering\arraybackslash}m{2.4cm}|>{\centering\arraybackslash}m{2.4cm}|}
\hline
\bf{Data Set} & \bf{Filtering-02} & \bf{HARD-03} & \bf{HARD-05} \\
\hline							
WA (Wikipedia Article) & 0.609 & 0.692 & 0.679\\
\hline
CT (Concept Title) & 0.647 & 0.629 & 0.663\\
\hline
AT (Anchor Texts) & 0.606 & 0.616 & 0.602\\
\hline
TD (Top Ranked Docs) & 0.661 & 0.597 & 0.531\\
\hline
RD (Related Docs) & 0.657 & 0.668 & 0.543\\
\hline
WA+CT & 0.654 & 0.726 & 0.693\\
\hline
WA+CT+AT & 0.670 & 0.750 & 0.702\\
\hline
WA+CT+AT+TD & 0.713 & 0.783 & 0.722\\
\hline
All & 0.735 & 0.785 & 0.722\\
\hline
\end{tabular}
\end{center}
\end{table}

\section{Conclusion}

We study a new user feedback mechanism based on Wikipedia concepts for interactive retrieval. To realize this mechanism, we propose several methods based on different sources of evidence for concept selection and document ranking. Our methods are evaluated on three TREC data sets along with a user study on Mechanical Turk. Experiment results show that user feedback on Wikipedia concepts can be very useful for document retrieval, and methods that combine all sources of evidence lead to the best performances for concept selection as well as document ranking.

In future work, we will continue our research in two directions. First, we will go beyond query expansion and study how the knowledge graph information of Wikipedia can be utilized for retrieval. Second, we will explore better methods for concept selection since this is a very important step. Active learning might be used in order to choose high-utility concept candidates. 

\bibliographystyle{splncs03}
\bibliography{ecir14}

\end{document}